1

# Finite Density Results for Wilson Fermions Using the Volume Method[*]

Walter Wilcox[a], Simeon Trendafilov[a] and E. Mendel[b]

[a]Department of Physics, Baylor University
Waco, TX 76798-7316 USA

[b]FB-8, Physik, Universität Oldenburg
26111 Oldenburg, Germany

Nonzero chemical potential studies with Wilson fermions should avoid the proliferation of flavor-equivalent nucleon states encountered with staggered formulation of fermions. However, conventional wisdom has been that finite baryon density calculations with Wilson fermions will be prohibitively expensive. We demonstrate that the volume method applied to Wilson fermions gives surprisingly stable results on a small number of configurations. It is pointed out that this method may be applied to any local or nonlocal gauge invariant quantity. Some illustrative results for $<\bar{\psi}\psi>$ and $<J>$ at various values of $\mu$ in a quenched lattice simulation are given.

## 1. INTRODUCTION

The behavior of quenched lattice QCD at finite chemical potential is poorly understood. In previous simulations using staggered fermions an early onset of the baryon phase transition occurs at $\mu \approx \frac{m_\pi}{2}$ opposed to the expected onset at $\mu \approx \frac{m_B}{3}$ where $m_B$ is the mass of the lightest baryon[1]. One possible explanation for the anomalous behavior of chemical potential is based on the stronger interaction than in nature of the 4 degenerate quark flavors present in the usual staggered formulation[2]. If this proposal is correct, then going to a single light flavor, as is the case for Wilson fermions, should substantially improve the situation.

Previously, calculations of thermal expectation values (THEVs) in the Wilson case have been hampered by the large amount of computer time necessary to invert the quark matrix starting at all inital space-time points. However, recent investigations of disconnected quark loops have been made possible by global algorithms which either use stochastic estimation techniques[3] or exploit gauge invariance[4] to estimate fermion matrix elements. We investigate here the use of the method in Ref. 4 (which we refer to as the "volume method") in calculating THEVs at finite chemical potential. We will see that this method gives numerically stable results on a surprisingly small number of gauge configurations for both local ($<\bar{\psi}\psi>$) and nonlocal ($<J>$) gauge invariant quantities.

## 2. VOLUME METHOD

The Wilson fermion action with chemical potential $\mu$ is given by

$$S_F = \sum_{\{i\},\{j\}} \bar{\psi}_{\{i\}} M_{\{i\}\{j\}} \psi_{\{j\}}, \quad (1)$$

where the collective indices $\{i\}$ and $\{j\}$ include space-time, color and Dirac indices and where

$$M_{\{i\}\{j\}} = \delta_{\{i\}\{j\}} - \kappa \sum_{k=1,2,3}[(1-\gamma_k)U_k(x)\delta_{x,y-a_k}$$
$$+(1+\gamma_k)U_k^\dagger(x-a_k)\delta_{x,y+a_k}]$$
$$-\kappa[e^{-\mu}(1-\gamma_4)U_4(x)\delta_{x,y-a_4}$$
$$+e^{\mu}(1+\gamma_4)U_4^\dagger(x-a_4)\delta_{x,y+a_4}]. \quad (2)$$

In a standard manner, one can calculate the THEV:

$$\sum_{\vec{x},t} <\bar{\psi}(\vec{x},t)\psi(\vec{x},t)> =$$
$$\sum_{\vec{x},t}\{Tr[M^{-1}(\vec{x},t;\vec{x},t)]\}_U. \quad (3)$$

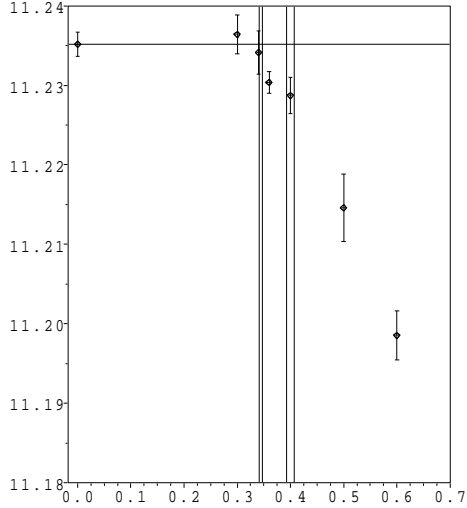 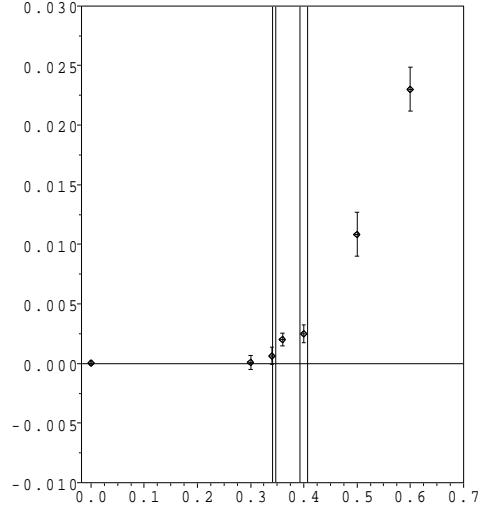

Figure 1. Values of $<\bar{\psi}\psi>$ as a function of $\mu$ at $\kappa=.148$. The horizontal line simply indicates the central value of $<\bar{\psi}\psi>$ at $\mu=0$. The two sets of vertical bars represent the error bars on measured values of $\frac{m_\pi}{2}$ and $\frac{m_\Delta}{3}$.

Figure 2. Values of the conserved charge density $<J>$ as a function of $\mu$ at $\kappa=.148$. The meanings of the vertical lines are the same as in Fig. 1.

(The notation $\{\cdots\}_U$ on the rhs indicates a gauge field average.) Because of the average over gauge configurations in Eq.(3), we may replace the point-to-point loop propagator $M^{-1}(\vec{x},t;\vec{x},t)$ with a version which is summed over all inital space-time positions. That is

$$\sum_{\vec{x},t}<\bar{\psi}(\vec{x},t)\psi(\vec{x},t)>=\sum_{\vec{x},t}\{Tr[M^{-1}(\vec{x},t)]\}_U \quad (4)$$

where

$$M^{-1}(\vec{x},t) \equiv \sum_{\vec{x}',t'} M^{-1}(\vec{x},t;\vec{x}',t'). \quad (5)$$

This is because

$$\{Tr[M^{-1}(\vec{x},t;\vec{x}',t')]\}_U = \{Tr[M^{-1}(\vec{x},t;\vec{x},t)]\}_U \delta_{\vec{x},\vec{x}'}\delta_{t,t'} \quad (6)$$

as a consequence of Elitzur's theorem[5]. The right hand side of Eq.(4) is simple to evaluate because it is just the trace of a quark propagator starting at all inital space-time points with unit weight.

Likewise, the volume method can be applied to the THEV of any gauge invariant quantity. The conserved and appropriately normalized nonlocal Wilson charge density operator at nonzero $\mu$ is given by

$$J(\vec{x},t) \equiv \kappa\{e^\mu\bar{\psi}(\vec{x},t+1)(1+\gamma_4)U_4^\dagger(\vec{x},t)\psi(\vec{x},t) \\ -e^{-\mu}\bar{\psi}(\vec{x},t)(1-\gamma_4)U_4(\vec{x},t)\psi(\vec{x},t+1)\}. \quad (7)$$

(The factors of $e^{\pm\mu}$ will remove any trivial $\mu$ dependence in $<J>$ from quarks which do not loop in the time direction.) We then have the THEV:

$$\sum_{\vec{x},t}<J(\vec{x},t)>= \\ \kappa\sum_{\vec{x},t}\{Tr[-e^\mu M^{-1}(\vec{x},t)(1+\gamma_4)U_4^\dagger(\vec{x},t) \\ +e^{-\mu}M^{-1}(\vec{x},t+1)(1-\gamma_4)U_4(\vec{x},t)]\}_U, \quad (8)$$

where we have again replaced the point-to-point space-time propagators by initial space-time summed ones, using Eq.(5). Since these expressions assume a gauge field average, an important numerical question is how well the gauge invariant signal in Eqs.(4) and (8) is projected out of the gauge variant noise. For this purpose we turn to the numerical simulations.

## 3. RESULTS

We present results on a small number of configurations in Figs. 1 and 2 for $<\bar{\psi}\psi>$ and $<J>$, respectively.

We used 5 gauge field configurations at each $\mu$ (.3, .34, .4, .5, and .6) except at $\mu = 0$ and .36 where 10 configurations were employed. Periodic boundary conditions were imposed on the quarks in the space direction and antiperiodic boundary conditions in the time direction, appropriate for finite temperature. The results come from $16^3 \times 24$ lattices at $\beta = 6.0$ and a relatively heavy quark mass hopping parameter of $\kappa = .148$. ($\kappa_c$ is approximately .1568.) In our conjugate gradient algorithm we found that it was very important to impose the convergence criterion directly on the quantity of interest. In this case we required that the relative change in $<\bar{\psi}\psi>$ over the previous 20 iterations to be less than $2 \times 10^{-5}$. (The number of iterations increased from about 60 at $\mu = 0$ to about 400 at $\mu = .6$.) Our raw, unsubtracted result for $<\bar{\psi}\psi>$ at $\mu = 0$ was $11.2352 \pm .0015$. This is in agreement with $11.23619 \pm .00031$ found with the Z2 stochastic method[6].

Fig. 1 represents the results of the $<\bar{\psi}\psi>$ measurements as a function of $\mu$ using Eq.(4). (Results in Figs. 1 and 2 have been normalized by the total number of space-time points.) The vertical lines on this figure (as well as the next) represent the error bar ranges on $\frac{m_\pi}{2}$ and $\frac{m_\Delta}{3}$ across 24 gauge field configurations ($m_\pi = .689 \pm .006$, $m_\Delta = 1.200 \pm .022$) [6] of which ours are a subset. (The $\Delta$ is the lowest mass single flavor three quark state.) As can be seen, there is a statistically significant departure from the zero chemical potential THEV when $\mu$ is below $\frac{m_\Delta}{3}$.

Fig. 2 shows the measurement of $<J>$ as a function of $\mu$ using Eq.(8). Again, there is a signicant departure from the vacuum value when $\mu$ is below $\frac{m_\Delta}{3}$. (At $\mu = 0$ we measure $<J> = .0004 \pm .0015$.) For $\mu > \frac{m_\Delta}{3}$, the values begin to fall ($<\bar{\psi}\psi>$) or rise ($<J>$) sharply.

As is evident from the figures, the volume method gives statistically stable results for both $<\bar{\psi}\psi>$ and the charge density, $<J>$. This is very encouraging considering the small number of configurations used. Thus, quenched finite density Wilson simulations are feasible without excessive amounts of computer time.

Since there is only one light Wilson fermion flavor, our results are compatible either with a shift to lower $\mu$ due to finite temperature[2] or a true early onset at $\mu = \frac{m_\pi}{2}$[7]. In order to clearly distinguish between these two interpretations, further measurements at lower quark masses, where the pion and delta mass scales are more widely separated, are planned.

## 4. ACKNOWLEDGEMENTS

This work was partially supported by the National Science Foundation under grant PHY-9401068 and the National Center for Supercomputing Applications and utilized the CM5 and CRAY-YMP systems at the University of Illinois at Urbana-Champaign. We are grateful to the authors of Ref. 3 who made their measurement of $m_\Delta$ and other data available. W.W. would like to acknowledge the Aspen Center for Physics where this investigation was begun.